\documentclass[reqno,12pt,a4paper]{amsart}

\usepackage{mathrsfs}
\usepackage{amssymb}
\usepackage{amsfonts}
\usepackage{latexsym}
\usepackage{amsthm}

\usepackage{graphicx}
\def\lb{\label}

\newcommand{\er}[1]{\textrm{(\ref{#1})}}

\begin{document}


\renewcommand{\theequation}{\arabic{equation}}
\theoremstyle{plain}
\newtheorem{theorem}{\bf Theorem}
\newtheorem{lemma}[theorem]{\bf Lemma}
\newtheorem{corollary}[theorem]{\bf Corollary}
\newtheorem{proposition}[theorem]{\bf Proposition}
\newtheorem{definition}[theorem]{\bf Definition}
\newtheorem{condition}[theorem]{\bf Condition}
\newtheorem{remark}[theorem]{\bf Remark}

\def\a{\alpha}  \def\cA{{\mathcal A}}     \def\bA{{\bf A}}  \def\mA{{\mathscr A}}
\def\b{\beta}   \def\cB{{\mathcal B}}     \def\bB{{\bf B}}  \def\mB{{\mathscr B}}
\def\g{\gamma}  \def\cC{{\mathcal C}}     \def\bC{{\bf C}}  \def\mC{{\mathscr C}}
\def\G{\Gamma}  \def\cD{{\mathcal D}}     \def\bD{{\bf D}}  \def\mD{{\mathscr D}}
\def\d{\delta}  \def\cE{{\mathcal E}}     \def\bE{{\bf E}}  \def\mE{{\mathscr E}}
\def\D{\Delta}  \def\cF{{\mathcal F}}     \def\bF{{\bf F}}  \def\mF{{\mathscr F}}
\def\c{\chi}    \def\cG{{\mathcal G}}     \def\bG{{\bf G}}  \def\mG{{\mathscr G}}
\def\z{\zeta}   \def\cH{{\mathcal H}}     \def\bH{{\bf H}}  \def\mH{{\mathscr H}}
\def\e{\eta}    \def\cI{{\mathcal I}}     \def\bI{{\bf I}}  \def\mI{{\mathscr I}}
\def\p{\psi}    \def\cJ{{\mathcal J}}     \def\bJ{{\bf J}}  \def\mJ{{\mathscr J}}
\def\vT{\Theta} \def\cK{{\mathcal K}}     \def\bK{{\bf K}}  \def\mK{{\mathscr K}}
\def\k{\kappa}  \def\cL{{\mathcal L}}     \def\bL{{\bf L}}  \def\mL{{\mathscr L}}
\def\l{\lambda} \def\cM{{\mathcal M}}     \def\bM{{\bf M}}  \def\mM{{\mathscr M}}
\def\L{\Lambda} \def\cN{{\mathcal N}}     \def\bN{{\bf N}}  \def\mN{{\mathscr N}}
\def\m{\mu}     \def\cO{{\mathcal O}}     \def\bO{{\bf O}}  \def\mO{{\mathscr O}}
\def\n{\nu}     \def\cP{{\mathcal P}}     \def\bP{{\bf P}}  \def\mP{{\mathscr P}}
\def\r{\rho}    \def\cQ{{\mathcal Q}}     \def\bQ{{\bf Q}}  \def\mQ{{\mathscr Q}}
\def\s{\sigma}  \def\cR{{\mathcal R}}     \def\bR{{\bf R}}  \def\mR{{\mathscr R}}
\def\S{\Sigma}  \def\cS{{\mathcal S}}     \def\bS{{\bf S}}  \def\mS{{\mathscr S}}
\def\t{\tau}    \def\cT{{\mathcal T}}     \def\bT{{\bf T}}  \def\mT{{\mathscr T}}
\def\f{\phi}    \def\cU{{\mathcal U}}     \def\bU{{\bf U}}  \def\mU{{\mathscr U}}
\def\F{\Phi}    \def\cV{{\mathcal V}}     \def\bV{{\bf V}}  \def\mV{{\mathscr V}}
\def\P{\Psi}    \def\cW{{\mathcal W}}     \def\bW{{\bf W}}  \def\mW{{\mathscr W}}
\def\o{\omega}  \def\cX{{\mathcal X}}     \def\bX{{\bf X}}  \def\mX{{\mathscr X}}
\def\x{\xi}     \def\cY{{\mathcal Y}}     \def\bY{{\bf Y}}  \def\mY{{\mathscr Y}}
\def\X{\Xi}     \def\cZ{{\mathcal Z}}     \def\bZ{{\bf Z}}  \def\mZ{{\mathscr Z}}
\def\O{\Omega}
\def\th{\theta}

\newcommand{\gA}{\mathfrak{A}}
\newcommand{\gB}{\mathfrak{B}}
\newcommand{\gC}{\mathfrak{C}}
\newcommand{\gD}{\mathfrak{D}}
\newcommand{\gE}{\mathfrak{E}}
\newcommand{\gF}{\mathfrak{F}}
\newcommand{\gG}{\mathfrak{G}}
\newcommand{\gH}{\mathfrak{H}}
\newcommand{\gI}{\mathfrak{I}}
\newcommand{\gJ}{\mathfrak{J}}
\newcommand{\gK}{\mathfrak{K}}
\newcommand{\gL}{\mathfrak{L}}
\newcommand{\gM}{\mathfrak{M}}
\newcommand{\gN}{\mathfrak{N}}
\newcommand{\gO}{\mathfrak{O}}
\newcommand{\gP}{\mathfrak{P}}
\newcommand{\gQ}{\mathfrak{Q}}
\newcommand{\gR}{\mathfrak{R}}
\newcommand{\gS}{\mathfrak{S}}
\newcommand{\gT}{\mathfrak{T}}
\newcommand{\gU}{\mathfrak{U}}
\newcommand{\gV}{\mathfrak{V}}
\newcommand{\gW}{\mathfrak{W}}
\newcommand{\gX}{\mathfrak{X}}
\newcommand{\gY}{\mathfrak{Y}}
\newcommand{\gZ}{\mathfrak{Z}}

\newcommand{\gm}{\mathfrak{m}}
\newcommand{\gn}{\mathfrak{n}}
\newcommand{\gf}{\mathfrak{f}}
\newcommand{\gh}{\mathfrak{h}}
\newcommand{\mg}{\mathfrak{g}}
\newcommand{\gb}{\mathfrak{b}}

\def\ve{\varepsilon}   \def\vt{\vartheta}    \def\vp{\varphi}    \def\vk{\varkappa}

\def\Z{{\mathbb Z}}    \def\R{{\mathbb R}}   \def\C{{\mathbb C}}    \def\K{{\mathbb K}}
\def\T{{\mathbb T}}    \def\N{{\mathbb N}}   \def\dD{{\mathbb D}}


\def\la{\leftarrow}              \def\ra{\rightarrow}            \def\Ra{\Rightarrow}
\def\ua{\uparrow}                \def\da{\downarrow}
\def\lra{\leftrightarrow}        \def\Lra{\Leftrightarrow}


\def\lt{\biggl}                  \def\rt{\biggr}
\def\ol{\overline}               \def\wt{\widetilde}
\def\no{\noindent}


\let\ge\geqslant                 \let\le\leqslant
\def\lan{\langle}                \def\ran{\rangle}
\def\/{\over}                    \def\iy{\infty}
\def\sm{\setminus}               \def\es{\emptyset}
\def\ss{\subset}                 \def\ts{\times}
\def\pa{\partial}                \def\os{\oplus}
\def\om{\ominus}                 \def\ev{\equiv}
\def\iint{\int\!\!\!\int}        \def\iintt{\mathop{\int\!\!\int\!\!\dots\!\!\int}\limits}
\def\el2{\ell^{\,2}}             \def\1{1\!\!1}
\def\sh{\sharp}
\def\wh{\widehat}
\def\bs{\backslash}

\def\sh{\mathop{\mathrm{sh}}\nolimits}
\def\Area{\mathop{\mathrm{Area}}\nolimits}
\def\arg{\mathop{\mathrm{arg}}\nolimits}
\def\const{\mathop{\mathrm{const}}\nolimits}
\def\det{\mathop{\mathrm{det}}\nolimits}
\def\diag{\mathop{\mathrm{diag}}\nolimits}
\def\diam{\mathop{\mathrm{diam}}\nolimits}
\def\dim{\mathop{\mathrm{dim}}\nolimits}
\def\dist{\mathop{\mathrm{dist}}\nolimits}
\def\Im{\mathop{\mathrm{Im}}\nolimits}
\def\Iso{\mathop{\mathrm{Iso}}\nolimits}
\def\Ker{\mathop{\mathrm{Ker}}\nolimits}
\def\Lip{\mathop{\mathrm{Lip}}\nolimits}
\def\rank{\mathop{\mathrm{rank}}\limits}
\def\Ran{\mathop{\mathrm{Ran}}\nolimits}
\def\Re{\mathop{\mathrm{Re}}\nolimits}
\def\Res{\mathop{\mathrm{Res}}\nolimits}
\def\res{\mathop{\mathrm{res}}\limits}
\def\sign{\mathop{\mathrm{sign}}\nolimits}
\def\span{\mathop{\mathrm{span}}\nolimits}
\def\supp{\mathop{\mathrm{supp}}\nolimits}
\def\Tr{\mathop{\mathrm{Tr}}\nolimits}
\def\BBox{\hspace{1mm}\vrule height6pt width5.5pt depth0pt \hspace{6pt}}
\def\as{\text{as}}
\def\all{\text{all}}
\def\where{\text{where}}
\def\Dom{\mathop{\mathrm{Dom}}\nolimits}


\newcommand\nh[2]{\widehat{#1}\vphantom{#1}^{(#2)}}
\def\dia{\diamond}

\def\Oplus{\bigoplus\nolimits}



\def\qqq{\qquad}
\def\qq{\quad}
\let\ge\geqslant
\let\le\leqslant
\let\geq\geqslant
\let\leq\leqslant
\newcommand{\ca}{\begin{cases}}
\newcommand{\ac}{\end{cases}}
\newcommand{\ma}{\begin{pmatrix}}
\newcommand{\am}{\end{pmatrix}}
\renewcommand{\[}{\begin{equation}}
\renewcommand{\]}{\end{equation}}
\def\eq{\begin{equation}}
\def\qe{\end{equation}}
\def\[{\begin{equation}}
\def\bu{\bullet}

\newcommand{\fr}{\frac}
\newcommand{\tf}{\tfrac}

\title[Inverse problem for 3-point Dirichlet problem]
{Inverse problem for 3-rd order operators under the 3-point
Dirichlet conditions}

\date{\today}
\author[Andrey Badanin]{Andrey Badanin}
\author[Evgeny Korotyaev]{Evgeny L. Korotyaev}
\address{Saint-Petersburg
State University, Universitetskaya nab. 7/9, St. Petersburg,
199034 Russia,
an.badanin@gmail.com,\  a.badanin@spbu.ru,\
korotyaev@gmail.com,\  e.korotyaev@spbu.ru}


\maketitle

\begin{abstract}
We solve an inverse problem for a third order differential operator
under the 3-point Dirichlet conditions.
The third-order operator is an $L$-operator
in the Lax pair for the good Boussinesq equation.
We construct the mapping from the set of the coefficients to the set of
spectral data. This mapping is an analytic bijection
on a neighborhood of the zero.
\end{abstract}

\section{Introduction}
Consider the {\it good Boussinesq equation}
$$
q_{xt}=-{1\/3}(p_{xxxx}+4(p^2)_{xx}), \qqq p_t=q_x,
$$
on the circle, which belongs to the class of integrable systems.
The $L$-operator in the Lax pair
for this equation is given by $L=\pa_x^3+p\pa_x+\pa_x p+q$.
The solutions to the Boussinesq equation are
parameterized by a {\it divisor}. The divisor
is a set of points on the 3-sheeted Riemann surface $\cR$ of the $L$-operator
whose projections coincide with the so-called {\it auxiliary spectrum},
which is the spectrum of the non-self-adjoint operator
\[
\lb{Hdpq}
\cL y=(y''+py)'+py'+qy,\qqq y(0)=y(1)=y(2)=0,
\]
with the 1-periodic coefficients $p,q$,
see \cite{McK81}. In this paper we consider the inverse problem of recovering
the coefficients $p,q$ from the auxiliary spectrum.

We consider the operator $\cL=\cL(\p)$ of the form \er{Hdpq},
acting on $L^2(0,2)$, where
$$
\p=(p,q)\in\gH,
$$
and $\gH$ is the real Hilbert space given by
$$
\gH=\cH_1\os\cH,\qq
\cH=\Big\{f\in L_\R^2(\T):\int_0^1f(x)dx=0\Big\},\qq \T=\R/\Z,\qq
\cH_1=\{f:f,f'\in\cH\}.
$$
It is equipped with the norm
$
\|\p\|_1^2=\|p'\|^2+\|q\|^2,
\|f\|^2=\int_0^1|f(x)|^2dx.
$
The spectrum of the operator $\cL$ is pure discrete.
In this paper, we consider the inverse problem of
recovering the coefficients $p, q$ from the auxiliary spectrum.

\section{Second order operators}
Recall the well known results about the Korteweg--de~Vries equation
$$
V_t=6VV_{x}-V_{xxx}
$$
on the circle.
$L$-operator in the Lax pair
for this equation is the Schr\"odinger operator $-\pa_x^2+V$ with
an 1-periodic potential $V$.
The solutions to the KdV equation are
parameterized by a divisor, which
is a set of points on the 2-sheeted Riemann surface of the $L$-operator
whose projections coincide with the auxiliary spectrum $\gm_1<\gm_2<...$
of the self-adjoint operator
$-y''+Vy$ under the Dirichlet boundary conditions $y(0)=y(1)=0$.
Projections of branch points of the Riemann surface coincide with
eigenvalues $\l_0^+<\l_1^-\le\l_1^+<...$ of the 2-periodic problem
for the Schr\"odinger operator. Each point $\gm_n,n\in\N$, belongs
to the interval $[\l_n^-,\l_n^+]$ and the corresponding point
of the divisor lies on one of two sheets of the Riemann surface.

Recall the inverse problem of recovering the potential
$V$ from the auxiliary spectrum.
Consider the operator $-y''+Vy$ with the potential
$V\in\cH$, acting in $L^2(0,1)$,
 under the Dirichlet boundary conditions $y(0)=y(1)=0$.
The spectrum consists of the simple eigenvalues $\gm_1<\gm_2<...$,
$\gm_n=(\pi n)^2+o(1)$, as $n\to+\iy$.
This eigenvalues are zeros of the entire function $\vp(1,\cdot)$,
where $\vp(x,\l)$ is the fundamental solution of the equation
\[
\lb{hilleq}
-y''+Vy=\l y,
\]
satisfying the initial conditions
$\vp|_{x=0}=0,\vp'|_{x=0}=1$.
Introduce the {\it norming constants}
$\gh_{sn}:\cH\to\R$ by
\[
\lb{ncschr}
\gh_{sn}(V)=2\pi n\log|\vp'(1,\gm_n(V))|,\qq\log 1=0,\qq n\in\N.
\]
Introduce the space $S$ of all real, strictly increasing
sequences $(s_1,s_2,...)$ such that $(s_n-(\pi n)^2)\in\ell^2$.
The following result holds true, see, e.g., \cite{PT87}.
\medskip

{\it
The mapping $V\to (\gm_n(V),\gh_{sn}(V))$
is a real analytic isomorphism between $\cH$ and $S\os\ell^2$.
}

\medskip

\no {\bf Remark.}
In order to recover the potential we need two sequences:

1) A sequence of the eigenvalues from $S$,

2) A sequence of the norming constants from $\ell^2$.

\medskip

The inverse problem of recovering the potential
$V$ from the periodic spectrum was solved by Korotyaev \cite{K99}.
Introduce the mapping $\mg:\cH\to \ell^2\os\ell^2$ from \cite{K99} by
\[
\lb{defwtg}
\mg(V)=(\mg_n(V))_{n\in\N},\qq V\in\cH,
\]
where
$$
\mg_n=(\mg_{cn},\mg_{sn})\in\R^2,
\qq |\mg_n|={\l_n^+-\l_n^-\/2},
$$
the vectors $(\mg_{cn},\mg_{sn})$ are given by
\[
\lb{defwtgcn}
\mg_{cn}={\l_n^++\l_n^-\/2}-\gm_n,\qq
\]
\[
\lb{defwtgsn}
\mg_{sn}=\Big|{(\l_n^+-\l_n^-)^2\/4}-\mg_{cn}^2\Big|^{1\/2}\sign \gh_{sn}.
\]
Introduce the Fourier transformations
$\F_{cn},\F_{sn}:\cH\to\R,n\in\N$, by
$$
f_{cn}=\F_{cn}f=\int_0^1f(x)\cos 2\pi nxdx,\ \
f_{sn}=\F_{sn}f=\int_0^1f(x)\sin 2\pi nxdx.
$$
and
$
\F_n=(\F_{cn},\F_{sn})_{n\in\N}.
$
The following results were proved by Korotyaev \cite{K99}.

\medskip

{\it The mapping $\mg$, given by
\er{defwtg}--\er{defwtgsn},
is a real analytic isomorphism between $\cH$ and $\ell^2\os\ell^2$.
Moreover, if $n\to+\iy$, then the functions $\mg_n$ satisfy
$$
\mg_n(V)=(V_{cn},V_{sn})+O(n^{-1}),
\qqq
{\pa\mg_n\/\pa V}=\F_n+O(n^{-1}).
$$
}

\medskip

\no {\bf Remark.}
\no 1. The potential may be recovered from three sequences:

1) Two sequences $\mg_{cn}$ and $\mg_{sn}$ from $\ell^2$,

2) A sequence of signs $\pm$.

\no 2. We construct the corresponding mapping for 3-rd order operators
in our paper \cite{BK24}.

\section{Eigenvalues}
Introduce the fundamental solutions $\vp_1, \vp_2, \vp_3$ to equation
\[
\lb{1b}
(y''+py)'+py'+q y=\l y,\qqq \l\in\C.
\]
satisfying the conditions $\vp_j^{[k-1]}|_{x=0}=\d_{jk},j,k=1,2,3$,
where $y^{[0]}=y$, $y^{[1]}=y'$, $y^{[2]}=y''+py$.
Each of the function $\vp_j(x,\cdot),x\in[0,1],j=1,2,3$, is entire
and real on $\R$.
The spectrum $\s(\cL)$ of the operator $\cL$ is pure discrete
and satisfies
\[
\lb{spec}
\s(\cL)=\{\l\in\C:\xi(\l)=0\},
\]
where $\xi$ is the entire function given by
\[
\lb{defsi}
\xi(\l)=\det\ma\vp_2(1,\l)&\vp_3(1,\l)\\
\vp_2(2,\l)&\vp_3(2,\l)\am.
\]

If $\p=0$, then all eigenvalues are simple, real, and have the form
\[
\lb{unpev}
\m_{n}^o=(z_n^o)^3,\qq z_n^o={2\pi n\/\sqrt3},\qq
n\in\Z_0:=\Z\sm\{0\}.
\]
It is obvious that the eigenvalues remain simple and real for small $p$ and $q$.
If the coefficients
$p,q$ are not small, then the eigenvalues may be non-real and multiple,
which greatly complicates the analysis, see \cite{BK21}.
For this reason, in this paper we consider the case of small coefficients
$p,q$.

Let $\ve>0$ be fixed and small enough.
Define the ball $\cB_\ve$ in $\gH$ by
$$
\cB_\ve=\{\p\in\gH:\|\p\|_1<\ve\},
$$
and the corresponding ball $\cB_{\C,\ve}$ in $\gH_\C$,
where $\gH_\C$ is the complexification of the real space $\gH$..

Let $\p\in\cB_\ve$. Then there is exactly one
simple real eigenvalue $\m_n$ of the operator $\cL$ in each domain
$\cD_n,n\in\Z_0$, where
\[
\lb{DomcD}
\cD_{n}=\{\l\in\C:|z-z_n^o|<1\},\qq
\cD_{-n}=\{\l\in\C:-\l\in\cD_n\},\qq n\ge 0,
\]
and there are not any other eigenvalues, here and below
$z=\l^{1\/3},\arg z\in(-{\pi\/3},{\pi\/3}]$.
Introduce the sequence
\[
\lb{defhcn}
h_{cn}=\m_n-\m_n^o, \qq n\in\Z_0.
\]
We formulate our results about the eigenvalues.

\begin{theorem}
\lb{Th3pram}
Each function $\mu_n,n\in\Z_0$, is analytic on the ball
$\cB_{\C,\ve}$. Moreover,
if $\p\in\cB_\ve$ and  $n\in\Z_0$, then
\[
\lb{as3pev}
\Big|h_{cn}+{\wh p_{sn}'\/\sqrt3}+\wh q_{cn}
-{\wh p_{cn}'\/3}+{\wh q_{sn}\/\sqrt3}\Big|
\le {C\|\p\|_1^2\/n},
\]
and the gradient $ {\pa\m_n(\p)\/\pa\p(t)} =\big({\pa\m_n(\p)\/\pa
p(t)},{\pa\m_n(\p)\/\pa q(t)}\big) $ satisfies
\[
\lb{aspamun}
\begin{aligned}
& \sup_{t\in[0,1]}
\Big|{\pa\m_n(\p)\/\pa p(t)}
-z_n^oa_n(t)\Big|
\le C\|\p\|_1,
\\
& \sup_{t\in[0,1]}\Big|{\pa\m_n(\p)\/\pa q(t)}+a_n(t)\Big|
\le {C\|\p\|_1\/|n|},
\end{aligned}
\]
for some $C>0$, where
\[
\lb{defa_n(t)}
a_n(t)=\cos(2\pi nt)+{\sin(2\pi nt)\/\sqrt3}.
\]
\end{theorem}

\section{Monodromy matrix and multipliers}
Define the {\it monodromy matrix} by
\[
\lb{defmm}
M(\l)=\big(\vp_j^{[k-1]}(1,\l)\big)_{j,k=1}^3.
\]
The matrix-valued function $M$ is entire.
The characteristic polynomial $D$ of the monodromy matrix is given by
\[
\lb{1c} D(\t,\l)
=\det(M(\l)-\t \1_{3}),\qq (\t,\l)\in\C^2.
\]
An eigenvalue  of $M$ is called a {\it multiplier}, it is a
zero of the polynomial $D(\cdot,\l)$.
The matrix $M$ has exactly $3$ (counting with
multiplicities) eigenvalues $\t_j,j=1,2,3$, which  satisfy
$
\t_1\t_2\t_3=1.
$
In particular, each $\t_j\ne 0$ for all $\l\in\C$.

The three functions $\t_j(\l),j=1,2,3$, constitute three branches
of a function that is analytic on the three-sheeted Riemann surface $\cR$,
mentioned above.
This surface is an invariant of the Boussinesq flow.
The surface $\cR$ has only algebraic singularities in $\C$.
These singularities are {\it ramifications} of the surface $\cR$.
The ramifications coincide with zeros of the {\it discriminant}
$\r$ of the polynomial $D$ given by
$$
\r=(\t_1-\t_2)^2(\t_1-\t_3)^2(\t_2-\t_3)^2.
$$
This is an entire function of $\l$.
Asymptotically at large $|\l|$
the surface $\cR$ approaches the Riemann surface of the function $\l^{1\/3}$.
In particular, it cannot break up into separate sheets.
The corresponding property of Riemann surfaces
is true for $n$-order operators with any $n\ge 2$, see \cite{BK12}.
Note that Riemann surfaces of matrix operators can be split
into separate parts consisting of one or several sheets, see \cite{BBK06},
\cite{CK06}, \cite{K08}, \cite{K10}.

In the unperturbed case $\p=0$ we have $\t_1=e^{e^{i{2\pi\/3}} z}$,
$\t_2=e^{e^{i{4\pi\/3}} z}$,
$\t_3=e^z$. The zeros of the function $\r$
coincide with $\m_n^o,n\in\Z$, and have multiplicity two.
If $\l\in\C\sm\{\m_{-n}^o,n\ge 0\}$, then the multiplier $\t_3$
is simple and positive on $\R_+$.

Therefore, if $\p\in\cB_\ve$ and
$\l\in\mD$,
where
\[
\lb{defmD3}
\mD=\C\sm\cup_{n\ge 0}\ol{\cD_{-n}},
\]
then the multiplier $\t_3(\l)$ is simple.
Moreover, in this case the function $\t_3$ is analytic on the domain $\mD$,
real for real $\l$, and $\t_3(\l)>0$ for all
$\l>1$.
Furthermore, there are exactly two real
zeros $r_n^\pm$ of the function $\r$ in each domain $\cD_n,n\in\Z$,
and there are not any other zeros. The 3-point eigenvalues $\m_n$ satisfy:
$\m_n\in[r_n^-,r_n^+]$ for all $n\in\Z_0$.

\section{Norming constants}
In order to define norming constants  we need to consider the transpose
operator $\wt\cL$ given by
\[
\lb{1btr}
\wt\cL y=-(\tilde y''+p\tilde y)'-p\tilde y'+q \tilde y=\l \tilde y
,\qqq \tilde y(0)=\tilde y(1)=\tilde y(2)=0,
\]
where $\p\in\cB_\ve$. In each
domain $\cD_n,n\in\Z_0$, there is exactly one real eigenvalue $\wt\m_n$ of
this operator.
Let $\wt y_n(x)$ be a corresponding eigenfunction such that $\wt y_n'(0)=1$.
The norming constants $h_{sn}, n\in \N$, are defined by
\[
\lb{defnf}
 h_{sn} =8(\pi n)^2\log |\wt y_n'(1)\t_3^{-{1\/2}}(\wt\m_n)|, \qq
\t_3^{1\/2}(\wt\m_n)>0.
\]
The proposed form \er{defnf} of norming constants
is based on the following considerations.
There exists a standard transformation that lowers the order of the equation
so that the 3rd order equation \er{1b} becomes
the Hill equation \er{hilleq} with an energy-dependent potential $V=V(E,\p)$
and $E={3 \/4}\l^{2\/3}$. We consider this transformation in \cite{BK24x}.
In particular, we prove there the following results.

{\it
Let $\p\in\cB_\ve$, let $n\in\N$, let $\gm_n$
be the eigenvalue of the Dirichlet problem for Eq.~\er{hilleq}
with the potential $V=V(E,\p)$, and let $\gh_{sn}$
be the corresponding norming constants,
 given by \er{ncschr}. Let
$h_{sn}$ be given by \er{defnf}.
Then
\[
\lb{relr3p2o}
\gm_n ={3\/4}\wt\m_n^{2\/3},\qq
\gh_{sn}={h_{sn}\/4\pi n}.
\]
}

In order to define the norming constants $h_{sn}$ for $n<0$ we use some
symmetries.
Introduce the vector valued function
$\p_*^-(x)=(p(1-x),-q(1-x))$, $x\in\R$.
Note that the operators $\cL(\p)$ and $-\cL(\p_*^-)$ are unitarily equivalent,
therefore,
the eigenvalues $\mu_n$ satisfy
$\m_{-n}(\p)=-\m_n(\p_*^-)$, and then
$h_{c,-n}(\p)=-h_{cn}(\p_*^-)$, for all $n\in\Z_0$.
We introduce the norming constants $h_{s,-n},n\in\N$, by
\[
\lb{def-nf}
h_{s,-n}(\p)=-h_{sn}(\p_*^-).
\]
The norming constants, defined by \er{defnf} and \er{def-nf}, satisfy the following
asymptotics.

\begin{theorem}
\lb{Thnf}
Each function $h_{sn},n\in\Z_0$, is analytic on the ball
$\cB_{\C,\ve}$. Moreover, if $\p\in\cB_\ve$, then
\[
\lb{asncr}
\Big|h_{sn}
-\Big({\wh p_{sn}'\/\sqrt3}+\wh q_{cn}
-\wh p_{cn}'+\sqrt3\wh q_{sn}\Big)\Big|\le{ C\|\p\|_1^2\/n},
\]
\[
\lb{asgrhcn}
\begin{aligned}
\sup_{t\in[0,1]}\Big|{\pa h_{sn}(\p)\/\pa p(t)}+z_n^ob_n(t)\Big|\le C\|\p\|_1,
\\
\sup_{t\in[0,1]}\Big|{\pa h_{sn}(\p)\/\pa q(t)}-b_n(t)\Big|
\le{C\|\p\|_1\/|n|},
\end{aligned}
\]
for all $n\in\Z_0$ and for some $C>0$, where
\[
\lb{defb_n(t)}
b_n(t)=\cos(2\pi nt)+\sqrt3\sin(2\pi nt).
\]
\end{theorem}

\section{Mapping}
Introduce the mapping $h:\p\to h(\p)$ on $\cB_\ve$ by
\[
\lb{defh}
h(\p)=(h_n(\p))_{n\in\Z_0},
\]
where the vector $h_n:\gH\to\R^2$ is given by
\[
\lb{defhn}
h_n=\big(h_{cn},h_{sn}\big),
\]
the first component $h_{cn}$ has the form \er{defhcn}
and the second one is defined by \er{defnf}.
The asymptotics \er{as3pev} and \er{asncr} show that
$h(\p)\in\ell^2\os\ell^2$, where
$$
\ell^2=\Big\{(a_n)_{n\in\Z_0}:\sum_{n\in\Z_0}|a_n|^2<\iy\Big\}.
$$
Introduce the linear isomorphism  $\cF:\gH\to\ell^2\os\ell^2$ given by
\[
\lb{defcF}
\cF\p=(\cF_n\p)_{n\in\Z_0}.
\]
where the linear transformations $\cF_n:\gH\to\C^2,n\in\Z_0$, have
the form
\[
\lb{defFhi} \cF_{n}\p=\ma-1&{1\/\sqrt3}\\1&-\sqrt3\am
\ma\F_{cn}&\F_{sn}\\-\F_{sn}&\F_{cn}\am\ma 0&1\\{1\/\sqrt3}&0\am
\ma p'\\q\am.
\]
We present our main result.

\begin{theorem}
\lb{ThNablagsn}

The mapping $h:\cB_{\C,\ve}\to h(\cB_{\C,\ve})$ is a real analytic
bijection between  $\cB_{\C,\ve}$ and $h(\cB_{\C,\ve})$ and satisfies
\[
\lb{estg-cF}
\|h(\p)-\cF\p\|\le C\|\p\|_1^2,
\]
for all $\p\in\cB_\ve$ and for some $C>0$.

\end{theorem}

\no {\bf Remark.}
1) Asymptotics of gradients of the mappings
$h_n$ are given by \er{aspamun} and \er{asgrhcn}.

\no 2) We think that similar results can be obtained not only in the
neighborhood of zero coefficients, but also in the neighborhood of
any coefficients for which all eigenvalues are simple. The situation
in the neighborhood of coefficients corresponding to multiple
eigenvalues is rather complicated and has not yet been studied.

\no 3) The inverse problem for the operator $iy'''+i(py)'+ipy'+qy$ on
the segment $[0,1]$ under the boundary conditions
$y(0)=y(1)=0,y'(0)=e^{i\phi}y(1),\phi\in[0,2\pi)$, is considered in
 Amour \cite{A01}.

\bigskip

\no\small {\bf Acknowledgments.}
Authors were supported by the RSF grant number
23-21-00023, internet page https://rscf.ru/project/23-21-00023/.

\end{document}